\documentclass[preprint]{revtex4-1}
\usepackage{amsmath,amssymb}
\usepackage{booktabs}
\usepackage[dvipdfm]{graphicx, color}
\usepackage{colordvi}
\usepackage{color}
\usepackage{float}
\usepackage{comment}
\usepackage[font=scriptsize]{caption}

\usepackage{lineno}
\date{}

\begin{document}
\title{Minimal Model for Stem-Cell Differentiation}
\author{Yusuke Goto}
\email[]{goto@complex.c.u-tokyo.ac.jp}
\author{Kunihiko Kaneko}
\email[]{kaneko@complex.c.u-tokyo.ac.jp} \affiliation{ Research
Center for Complex Systems Biology, Graduate School of Arts and
Sciences, The University of Tokyo, 3-8-1 Komaba, Meguro-ku, Tokyo
153-8902, Japan}

\begin{abstract}
To explain the differentiation of stem cells in terms of dynamical
systems theory, models of interacting cells with intracellular
protein expression dynamics are analyzed and simulated. Simulations
were carried out for all possible protein expression networks
consisting of two genes under cell--cell interactions mediated by
the diffusion of a protein. Networks that show cell differentiation
are extracted and two forms of symmetric differentiation based on
Turing's mechanism and asymmetric differentiation are identified. In
the latter network, the intracellular protein levels show
oscillatory dynamics at a single-cell level, while cell-to-cell
synchronicity of the oscillation is lost with an increase in the
number of cells. Differentiation to a fixed-point type behavior
follows with a further increase in the number of cells. The cell
type with oscillatory dynamics corresponds to a stem cell that can
both proliferate and differentiate, while the latter fixed-point
type only proliferates. This differentiation is analyzed as a
saddle-node bifurcation on an invariant circle, while the number
ratio of each cell type is shown to be robust against perturbations
due to self-consistent determination of the effective bifurcation
parameter as a result of the cell--cell interaction. Complex cell
differentiation is designed by combing these simple two-gene
networks. The generality of the present differentiation mechanism,
as well as its biological relevance, is discussed.
\end{abstract}
%\keywords{???}
\maketitle

\section{Introduction}

The differentiation of multipotent stem cells into lineage-specific
cells is an important process in developmental biology, for
which an understanding in terms of dynamical systems theory is
desired. Stem cells can both proliferate (i.e., reproduce
themselves) and differentiate into other cell types
\cite{Lenza,Potten,Melton,Slack}. While the former indicates that
the cellular state is stable in retaining its composition, the
latter implies that the original cellular state is also unstable in
that it moves toward a state with different compositions. How a stem
cell simultaneously supports these two conflicting features is an
important question.

Dynamical system approaches to cell differentiation have been
developed by considering that abundances of cellular components are
changed through intracellular reactions
\cite{Waddington,Newman,Goodwin,Glass,Kauffman,Kauffman2,Huang}.
Protein expression dynamics consist of mutual activation and
inhibition, and the concentration of each protein changes over time,
while its composition determines the cellular state. A dynamical
system in the state space that consists of each protein expression
level can then be considered. Based on this picture, it is natural
to assign an attractor to each cell type, which allows the
robustness of each cell type to be explained as the stability of an
attractor against noise. If the system has multiple attractors, each
differentiated cell type corresponds to each attractor.

Although this attractor picture is important for understanding the
robustness of each cellular state and how distinct cell types are
formed, answers to the following questions remain elusive: (i) Which
attractor describes the two conflicting functions in stem cells,
i.e., proliferation and differentiation? (ii) How are initial
conditions for different attracting states selected through the
course of development? (iii) How is the stability of the
developmental course, i.e., the robustness in the timing of cell
differentiation, and the number distribution of each cell explained,
which possibly includes regulation of the cell--cell interactions?

To address these questions, coupled dynamical systems that include
intracellular dynamics of protein abundances, cell--cell
interaction, and an increase in the cellular number by cell division
were developed
\cite{KK-Yomo94,KK-Yomo97,KK-Yomo99,Furusawa-KK98,Furusawa-KK01}.
Cells with oscillatory intracellular dynamics of protein
concentrations were shown to differentiate to a novel type with an
increase in the cell number, and a cell model whose protein
expressions are regulated by a gene regulatory network (GRN) with 5
genes has recently been studied \cite{SFK}. From extensive
simulations, GRNs that include ``stem cells'' that can undergo both
proliferation and differentiation were selected. The cells always
exhibited oscillatory expression dynamics as a single-cell level,
and after divisions, synchronization of the oscillations among cells
was lost. With further cell division, differentiation to a cell type
that loses the oscillatory expression followed. Furthermore, it is
interesting to note that recent measurements of protein expression
dynamics in embryonic stem cells support this oscillation scenario
\cite{Kobayashi}. The importance of intracellular oscillation to
cell differentiation is now being recognized in the developmental
biology community \cite{Furusawa-KK12}.

To analyze this differentiation mechanism in terms of dynamical
systems theory, however, it would be useful to adopt a simpler
system consisting of fewer genes. Here, we study such a minimal
system, i.e., dynamical systems consisting of expression levels of
only two genes (proteins). After introducing the model in \S II, we
describe the results in \S III. By simulating all possible
regulation networks consisting of two proteins with various
parameters, we extract a minimal system that allows for stem-cell
differentiation. In such a system, the protein expression levels
oscillate in time as a single-cell level. With an increase in the
cell number, synchronization in the oscillation among cells is lost
due to cell--cell interactions and some cells then differentiate to
fall into fixed-point dynamics losing the oscillation. This process
is analyzed using bifurcation theory and explained in terms of a
saddle-node on an invariant circle (SNIC) bifurcation. We also show
that the number of each cell type after development is robust
against noise. The extracted two-gene network leading to the SNIC is
shown to provide a universal motif for asymmetric differentiation
from stem cells, while complex hierarchical differentiation is
designed by simply combining the extracted two-gene network in
parallel or in sequence. The relevance of the present results to
dynamical systems and biological cell differentiation is discussed
briefly in \S IV.

\section{Model}

\subsection{Intracellular protein expression dynamics}

Here we describe our cell differentiation model under cell-to-cell
interactions. Cell states are represented by protein expression
levels of two genes $x$ and $y$. These $x$ and $y$ genes can
regulate the protein expression level of both itself and the other
gene. We consider the dynamics of the expression levels (protein
concentrations) of the two proteins, denoted as $x_{i}(t)$ and
$y_{i}(t)$, of the $i$-th cell at time $t$. Following earlier
studies, we eliminate the mRNA concentration synthesized from each
gene and obtain the dynamical system only of the protein expression
levels. The dynamics of the expression levels at a single cell are
described as
\begin{equation}
\begin{split}
\frac{dx_{i}(t)}{dt}=\frac{1}{1+exp\{ -\beta( J_{xx}x_{i}(t)+J_{xy}y_{i}(t)-g_{x})\} }-x_{i}(t)\\
\frac{dy_{i}(t)}{dt}=\frac{1}{1+exp\{ -\beta( J_{yx}x_{i}(t)+J_{yy}y_{i}(t)-g_{y})\} }-y_{i}(t).\\
\end{split}
\end{equation}
\\
The $2\times2$ matrix $J_{m\ell}$ ($m,\ell=x,y$) gives the
transcriptional regulation from protein $\ell$ to $m$, where
$J_{m\ell}=1$ if the gene $\ell$ activates the expression of $m$,
$-1$ if it inhibits the expression, and 0 if there is no regulation.
A sigmoid function of the form $exp(-\beta (z-g))$ is adopted
\cite{gene-net} to represent the on--off type expression with
$\beta$ as the sensitivity parameter, which roughly corresponds to
the Hill coefficient in terms of cell biology, where for $\beta
\rightarrow \infty$ the function approaches the step function. The
parameters $g_{x}$ and $g_{y}$ give the threshold values of the
expressions of the $x$ and $y$ genes, respectively.

\subsection{Cell--cell interaction}
The cells interact with each other and this interaction is mediated
by some signal. We assume here that the interaction is mediated
directly or indirectly by one of the proteins, which we take to be
$y$. In its simplest form, we consider only the interaction by the
diffusion of the $y$ protein. By assuming that the diffusion is fast
and by discarding spatial inhomogeneity over the cells, we employ
global coupling, i.e., an all-to-all mean-field interaction. Thus,
the overall gene expression dynamics obey the following equation:
\begin{equation}
\begin{split}
\frac{dx_{i}(t)}{dt}=\frac{1}{1+exp\{ -\beta( J_{xx}x_{i}(t)+J_{xy}y_{i}(t)-g_{x})\} }-x_{i}(t)\\
\frac{dy_{i}(t)}{dt}=\frac{1}{1+exp\{ -\beta( J_{yx}x_{i}(t)+J_{yy}y_{i}(t)-g_{y})\} }-y_{i}(t)+\\
\frac{D}{N(t)}(\sum_{k=1}^{N}y_{k}(t)-y_{i}(t)),\\
\end{split}
\end{equation}
\\
where $D$ is the diffusion constant, and $N(t)$ is the total number
of cells at time $t$.

\subsection{Cell division}

To examine the cell differentiation process through development, the
number of cells $N(t)$ is increased over time by cell division.
Starting with a single cell, this cell is divided into two cells
that have almost the same protein concentrations at a certain time.
The concentrations $x_i(t)$ and $y_i(t)$ are slightly perturbed by
cell division, leading to cell differences that obey a Gaussian
distribution with the variance $\sigma$, which is set sufficiently
small. Here, the behavior to be discussed (i.e., whether cell
differentiation appears) is independent of the noise level. Noise is
included mainly to remove synchronization over cells, which is
unstable but preserved due to numerical computation with a finite
bit. The cells are simply divided into two after every time span
$t_d$.

\subsection{Model parameters and criterion for differentiation}

To numerically investigate this model, we set the sensitivity
parameter as $\beta=40$ and the division noise level as
$\sigma=10^{-3}$. The results do not depend on these specific
values, as long as the former is sufficiently large (e.g., $\beta
>5$) and the latter is not too large (e.g., $\sigma < 0.1$). The
division time $t_d$ is chosen to be 25, but this specific value also
does not affect the results.

We scanned the other parameters for all possible choices of
$J_{m\ell}=\pm1,0$ ($m,\ell = x $ or $y$) to examine the possibility
of cell differentiation. We also checked all possible configurations
of $J_{m\ell}$ as long as the matrix includes at least one
off-diagonal component (i.e., as a minimum, the interaction between
$x$ and $y$ exists). This gives a total of 36 configuration. The
parameters $g_{x,y}$ were varied from $-1$ to 1 in increments of
0.05, and $D$ was chosen as either 0.2 or 1.0. Thus, the model in
Eq.(2) was simulated for a total of $36\times 41\times 41 \times
2=121032$ cases.

To examine if cells differentiated, we increased the cell population
to $N=32$. For each cell, we computed the temporal average of the
gene expression level for a sufficiently long time after discarding
the transient time. If the difference between the maximum and
minimum averaged gene expression levels ($x_i(t)$, $y_i(t)$) over
the cells was larger than 0.1, then we concluded that
differentiation had occurred.

\section{Results}

\subsection{Differentiation classification}

\begin{center}
\begin{tabular}{cccc} \hline
Type & Behavior  & Network & Parameter\\ \hline
Turing & fixed point\;\;$\rightarrow$ fixed point  & 2/36 & 6/121032\\
Oscillation death &  oscillation\;\;$\rightarrow$ fixed point  &2/36 &3/121032\\
Asymmetric differentiation  & oscillation$\rightarrow$  & &\\
with oscillation & oscillation  & 2/36 &9/121032\\ \hline
\end{tabular}
\end{center}

The simulation results revealed cell differentiation cases that can
be classified into the following three types: 1) Turing, 2)
Oscillation death, and 3) Asymmetric differentiation with remnant
oscillations. In all cases, the single-cell dynamics had only a
single attractor. As the cell number increases, the cells take two
distinct states. We first describe these three types and show that
the mechanisms of the first two types are already well known. We
thus focus on the third type, which is the most relevant to
differentiation of stem cells.

\subsubsection{Turing type (fixed point$\rightarrow$fixed point)}

\begin{figure}[H]
\begin{center}
\includegraphics[width= 100mm]{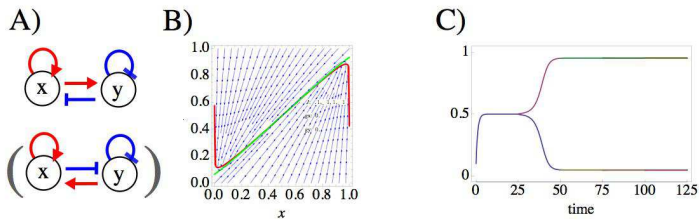}
\caption{(a) Upper: Network structure %and  parameter values
of a model that shows Turing-type differentiation; fixed point
$\rightarrow$ fixed point. Lower: The conjugate network form. The
parameter values that show this differentiation are
$(D,g_{x},g_{y})=(0.2,0,0),(1,0,0),(1,0.05,0.05),(1,-0.05,-0.05)$
for the upper network and
$(D,g_{x},g_{y})=(0.2,1,-1),(1,1,-1),(1,0.95,-0.95)$ for the lower
network. (The flow in the state space and the nullcline is simply a
mirror symmetry with regard to $x=y$ in the model here and is not
shown). (b) The single-cell nullcline of the model. (c) Time series
of the protein expression $x_i(t)$ as the cell number increases from
$i=1$ to 32 through cell division every 25 time units; (b) and (c)
are drawn for the upper network structure with the parameter values
$(D,g_{x},g_{y})=(0.2,0,0)$. \label{Fig221}}
\end{center}

\end{figure}

The single-cell dynamics in this case has a unique stable fixed
point. As the cell number increases, the cell population splits into
two groups, each taking different fixed-point values of $x$ and $y$
and a higher value of $x$ or $y$. The numbers in each population are
equal, i.e., 16 cells each for $N=32$. One gene expression network
type is shown in Fig. \ref{Fig221}, where one gene activates the
expression of itself and the other, while the other gene, which is
diffusible, inhibits the expression of itself and the other. This
mechanism is explained well by the classic Turing pattern
\cite{Turing} by replacing the spatially local diffusive interaction
therein with global coupling. Indeed, Turing's seminal paper
included the present case, as he discussed the case with $N=2$ (see
also \cite{Sano}). The conjugate network is also shown in Fig.
\ref{Fig221} (in parentheses), where the differentiation mechanism
is understood in the same way.

\subsubsection{Oscillation death (oscillation $\rightarrow$ fixed point)}

\begin{figure}[H]
\begin{center}
\includegraphics[width=100mm]{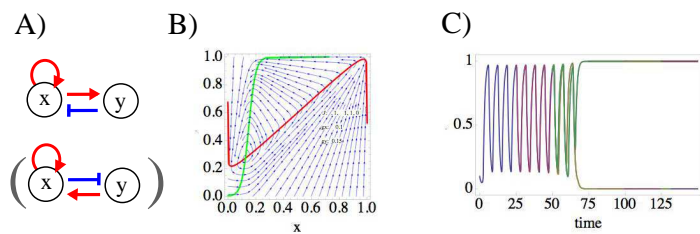}
\caption{(a) Upper: Network structure of a model that shows
differentiation through oscillation death; oscillation $\rightarrow$
fixed point. Lower: The conjugate network form. The parameter values
that show this differentiation are
$(D,g_{x},g_{y})=(1,-0.1,0.15),(1,0.1,085)$ for the upper network
and $(D,g_{x},g_{y})=(1,0.9,-0.15)$ for the lower network. (The flow
in the state space and the nullcline is simply a mirror symmetry
with regard to $x=y$ in the model here and is not shown). (b) The
single-cell nullcline of the model. (c) Time series of the protein
expression $x_i(t)$ as the cell number increases from $i=1$ to 32
through cell division every 25 time units; (b) and (c) are drawn for
the upper network structure with the parameter values
$(D,g_{x},g_{y})=(1,-0.1,0.15)$. } \label{Fig222}
\end{center}
\end{figure}

The single-cell dynamics for this type of differentiation has a
limit-cycle attractor. With an increase in cell number, the cell
population again splits into two groups of equal number, each of
which shows distinct fixed points through the cell-to-cell
interaction, in the same manner as the Turing-pattern case. This
loss of oscillation is known as oscillation death
\cite{Turing,Prigogine,oscillation-death,Koseska}. Two forms of the
networks that show this behavior are shown in Fig. \ref{Fig222}.
The mechanism of this type of differentiation will be explained later, 
as it is common to that of asymmetric differentiation in the next section.

\subsubsection{Asymmetric differentiation with remnant oscillation (oscillation $\rightarrow$ oscillation transition) }

In this class, the attractor of a single cell is again a limit
cycle, but as the number of cells is increased, synchronized
oscillation over cells is unstable under the cell--cell interaction,
and protein-expression oscillations are desynchronized between some
cells \cite{GCM}. After this desynchronization, some cells leave the
original limit-cycle trajectory and enter an oscillatory state with
a tiny amplitude, while the other cells remain in the original limit
cycle (with a slight modification) \cite{Meinhardt}. The states
mutually stabilize their existence, e.g., if all the cells in one
state were removed, some of the remaining cells will make a
transition to the other state. Thus, two distinct oscillatory states
coexist independently of the initial condition of the cells.

\begin{figure}[H]
\begin{center}
\includegraphics[width=100mm]{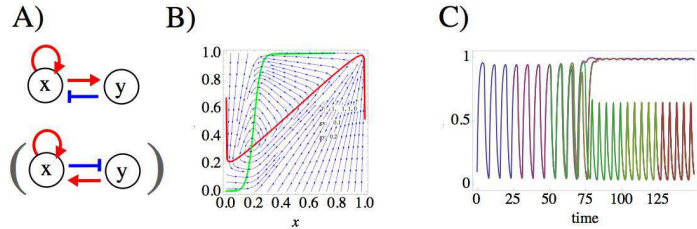}
\caption{ (a) Upper: Network structure of a model that shows
asymmetric differentiation with remnant oscillation; oscillation
$\rightarrow$ oscillation transition. Lower: The conjugate network
form. The parameter values that show this differentiation are
$(D,g_{x},g_{y})=(0.2,-0.1,0.15),(0.2,-0.1,0.2),(0.2,-0.1,-0.25)$
for the upper network and
$(D,g_{x},g_{y})=(0.2,0.1,0.75),(0.2,0.1,0.8),(0.2,0.1,0.85)$ for
the lower network. (The flow in the state space and the nullcline is
simply a mirror symmetry with regard to $x=y$ in the model here and
is not shown). (b) The single-cell nullcline of the model. (c) Time
series of the protein expression $x_i(t)$ as the cell number
increases from $i=1$ to 32 through cell division every 25 time
units; (b) and (c) are drawn for the upper network structure with
the parameter values $(D,g_{x},g_{y})=(0.2,-0.1,0.2)$. }
\label{Fig223}
\end{center}
\end{figure}

This class of behavior is observed for the two networks, shown in
Fig. \ref{Fig223}, over 9 sets of parameter values. The parameter
interval in which the present behavior is observed is relatively
small compared with the previous two cases.

\subsection{Mechanism of oscillatory differentiation}

We note that none of the cells in the first two classes have the
potential to both proliferate (reproduce the same type by division)
and differentiate (switch to a cell type with distinct behavior).
Furthermore, the first two mechanisms are already well understood
based on Turing's study, while the latter mechanism, which
corresponds to the cell-differentiation mechanism in our earlier
studies, is not fully understood in terms of dynamical systems
theory. Hence, we focus here on the third class and explain the
mechanism of the differentiation into two states.

\begin{figure}[H]
\begin{center}
\includegraphics[width=100mm]{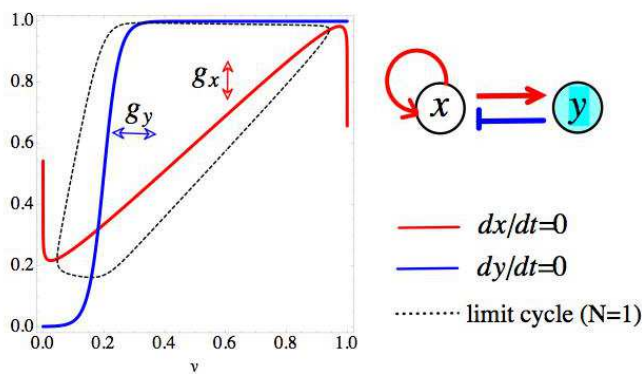}
\caption{The nullcline and limit-cycle attractor of the single-cell
dynamics for the GRN in Fig. \ref{Fig223}. The network is redrawn
for reference, where the parameter values are
$J_{xx}=1,J_{xy}=-1,J_{yx}=1,J_{yy}=0,g_x=-0.1,g_y=0.2$. The two
nullclines for $dx/dt=0$ and $dy/dt=0$ are drawn in red and blue,
respectively. The limit-cycle attractor is drawn as the dashed line.
The two nullclines come close but do not intersect, which is an
important property for differentiation to occur by cell--cell
interaction. Indeed, the nullclines and the limit cycle are arranged
in the same manner for the conjugate network (shown in parentheses
of Fig. \ref{Fig223}a). } \label{Fig231}
\end{center}
\end{figure}

The differentiation is based on two stages: desynchronization of the
oscillations and SNIC bifurcation induced by cell-to-cell
interactions \cite{SNIC,SNIC2}. Examining the nullcline of the
single-cell dynamics, shown in Fig. \ref{Fig231}, we see that while
the two nullclines come close, they do not intersect. Thus, there
are no fixed points, and the limit-cycle attractor exists as a
single-cell level, as already mentioned. (For the following
numerical simulations, we have used the parameter values $g_x=-0.1$,
$g_y=0.2$, and $D=0.14$).

\begin{figure}[H]
\begin{center}
\includegraphics[width=100mm]{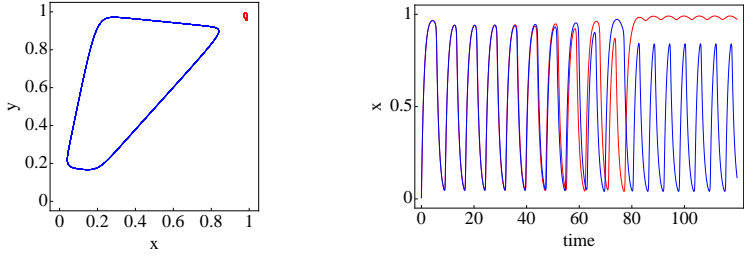}
\caption{Dynamics of the model in Figs. \ref{Fig223} and
\ref{Fig231} with $D=0.14$. (a) Locus of the orbit of the two-cell
dynamics of $(x_(t),y_1(t))$ and $(x_2(t),y_2(t))$. Blue: cell type
whose state is close to the original limit cycle. Red:
differentiated type with loss of autonomous oscillation. (b) The
time series of $x_i(t)$ for the two-cell dynamics ($i=1,2$). The
oscillations of the two cells are first desynchronized and at around
$t\sim 80$, they split into two distinct steady states. }
\label{Fig232}
\end{center}
\end{figure}

\subsubsection{Desynchronization}

Consider a coupled system with cell-to-cell interaction, where each
cell shows a limit-cycle oscillation as shown above. As is known,
the oscillations of the cells lose synchronization for a certain
range of the interaction and parameter values. This
desynchronization occurs in the present case, as was numerically
confirmed by computing the stability exponent for the
synchronization. Indeed, the magnitude of the tangential vector
representing the deviation between the two cells $\delta {\bf
X}$=($\delta x,\delta y)$  increases exponentially over time.

\subsubsection{Differentiation}

After desynchronization, the cell-to-cell interaction leads to
differentiation with an increase in the cell number. Here, the
cell--cell interaction term that occurs through diffusion in our
model is represented by $\alpha(t) \equiv
D(\frac{1}{N}\sum_{k=1}^{N}y_k-y_i)=D(\overline{y}-y_i) $, so that
the dynamics are written as
\begin{equation}
\begin{split}
\frac{dx(t)}{dt}=\frac{1}{1+exp\{ -\beta( x(t) - y(t)-g_{x})\} }-x(t)\\
\frac{dy(t)}{dt}=\frac{1}{1+exp\{ -\beta( x(t)-g_{y})\} }-y(t)+\alpha. \\
\end{split}
\end{equation}
If the oscillations are synchronized over cells, then $\alpha\equiv
0$, and the equation is reduced to single-cell dynamics. Due to the
desynchronization, however, $\alpha$ is nonzero and functions as a
time-dependent bifurcation parameter. We first examine the
bifurcation of the single-cell dynamics against changes in the
constant parameter $\alpha$ and then discuss the time dependence of
$\alpha$.

\begin{figure}[H]
\begin{center}
\includegraphics[width=100mm]{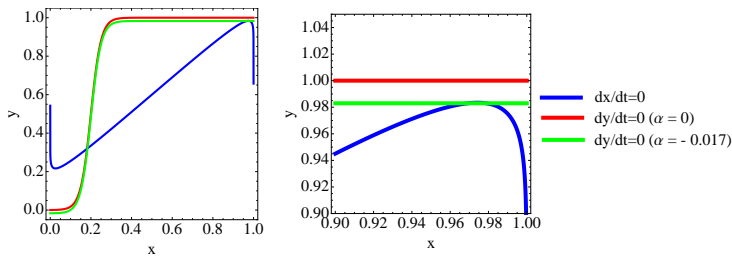}
\caption{(a) Change in the nullcline of $dy/dt=0$ as $\alpha$ in Eq.
(3) changes. As $\alpha$ is decreased, the nullcline $dy/dt=0$ moves
downwards (from the red to green line) and touches the nullcline of
$dx/dt=0$ (blue), when $\alpha=-0.017$. The SNIC then follows.
Details are shown in (b). } \label{Fig231no1}
\end{center}
\end{figure}

\begin{figure}[H]
\begin{center}
\includegraphics[width=80mm]{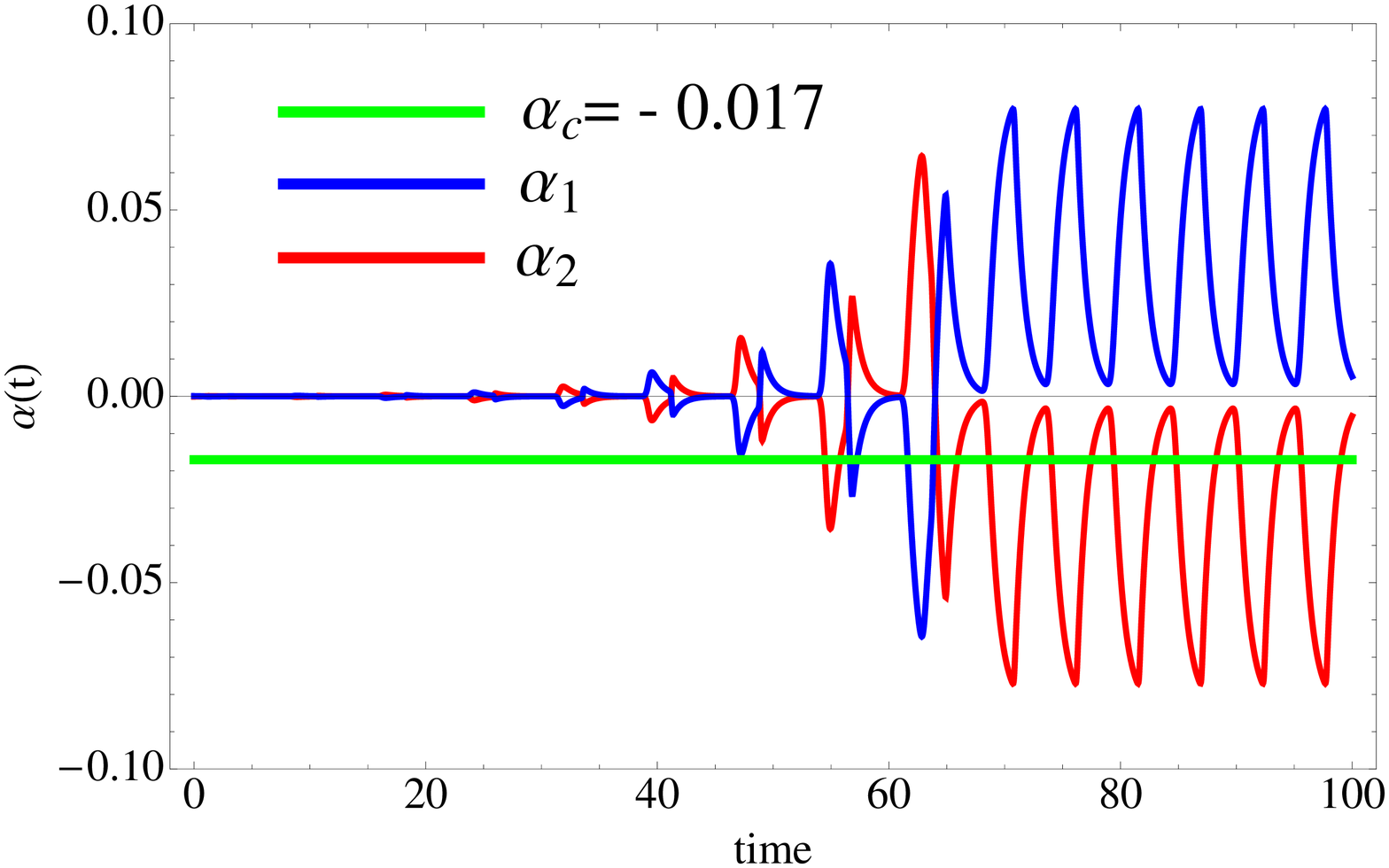}
\caption{Time series of the diffusion terms $D(y_2(t)-y_1(t))$
(blue) and $D(y_1(t)-y_2(t))$ (red) of the two cells. The cell--cell
interaction induces desynchronization, and the difference between
the cells is amplified. One cell then stays in the region of the
fixed point. For reference, the value $\alpha_c =-0.017$ is plotted.
Note that $D(y_1(t)-y_2(t))$ remains below $\alpha_c$ for most of
the time, while $D(y_2(t)-y_1(t))$ remains above it. }
\label{Fig231no2}
\end{center}
\end{figure}

The nullclines of Eq. (2) are shown in Fig. \ref{Fig231no1}. As
$\alpha$ is changed, only the nullcline $dy/dt=0$ moves vertically
downward with a decrease in $\alpha$ without a change in form; there
is no effect on the nullcline $dx/dt=0$. The two nullclines
intersect with a slight decrease in $\alpha$, so that the SNIC
\cite{SNIC,SNIC2} appears, and the limit-cycle attractor is replaced
by a stable fixed point. (In the example in Fig. \ref{Fig231no1},
this bifurcation occurs at $\alpha_c \approx -0.017 $).
%%%%%%%%%%%%%%%%%%%%%%%%%%%%%%%%%%%%%325-added
In the case in which $\overline{y}$, i.e., the average of $y$ over
$\alpha$, is represented as a constant value that can deviate from
$y_i(t)$, the nullcline for $dy/dt=0$ is represented by
$y=y_c=1/\{(1+exp\{- \beta (x-g_{y}) \})(1+D)\}- D\overline{y}$ and it decreases if $D$ increases.
 Recalling that the $x$ nullcline $y = x+\frac{1}{\beta}log(1-1/x)-g_{x} \approx 1$ in the region near$ x=1$(see figure),
 $x$ and $y$ nullcline intersect and a fixed point appears via the SNIC, if $D$ is large enough.
%%%%%%%%%%%%%%%%%%%%%%%%%%%%%%%%%%%%%%%%%%%%
In the differentiation of two cells that originates in
desynchronization, as shown in Fig. \ref{Fig232}, one cell has a
larger $y_i$ value, and the $\alpha$ $(<0)$ value of that cell is
smaller and can be smaller than $\alpha_c$. The nullcline $dy/dt=0$
for such a cell then intersects the nullcline $dx/dt=0$. In
contrast, the other cell remains close to the original limit cycle.
The effective $\alpha_{1,2}$ estimated from the diffusion term,
shown in Fig. \ref{Fig231no2}, demonstrates that $\alpha_2$ remains
in the region of the fixed point($\alpha<\alpha_c$), which supports
this argument. Thus, for a certain parameter region, the two types
of behavior are differentiated: one is close to the original limit
cycle and the other is close to the fixed point generated by the
SNIC.


Since the variables $x$ and $y$ oscillate in time, this argument
assuming a constant $\alpha$ is insufficient. Still, on average, the
trajectory of the differentiated cells that take a larger $y$ value
remains above the nullcline $dy/dt=0$ in the state space $(x,y)$.
Although the trajectory stays around the intersection of the two
nullclines, the state of this cell is not completely a fixed point
due to the interaction term with the other cell: since the other
cell type keeps the oscillatory dynamics close to the original limit
cycle, the differentiated cell type is driven by this oscillatory
dynamics so that it shows an oscillation with a tiny amplitude.  The
original and differentiated cell types are distinguished by the
ability for autonomous oscillation.
%%%%%%%%%%%%%%%%%%%%%%%%%%%%%%%%%%%%%%%%%%%326-added
Note that oscillation death(case2) is understood similary as the asymmetric differentiation.
In thats case,
SNIC bifurcation occur at two points around $(x,y) = (0,0)$ and $(1,1)$ on $x$-$y$ plane,
 and the limit cycle is replaced by the two fixed points as a result.

\subsubsection{Parameter dependence}

\begin{figure}[H]
\begin{center}
\includegraphics[width=125mm]{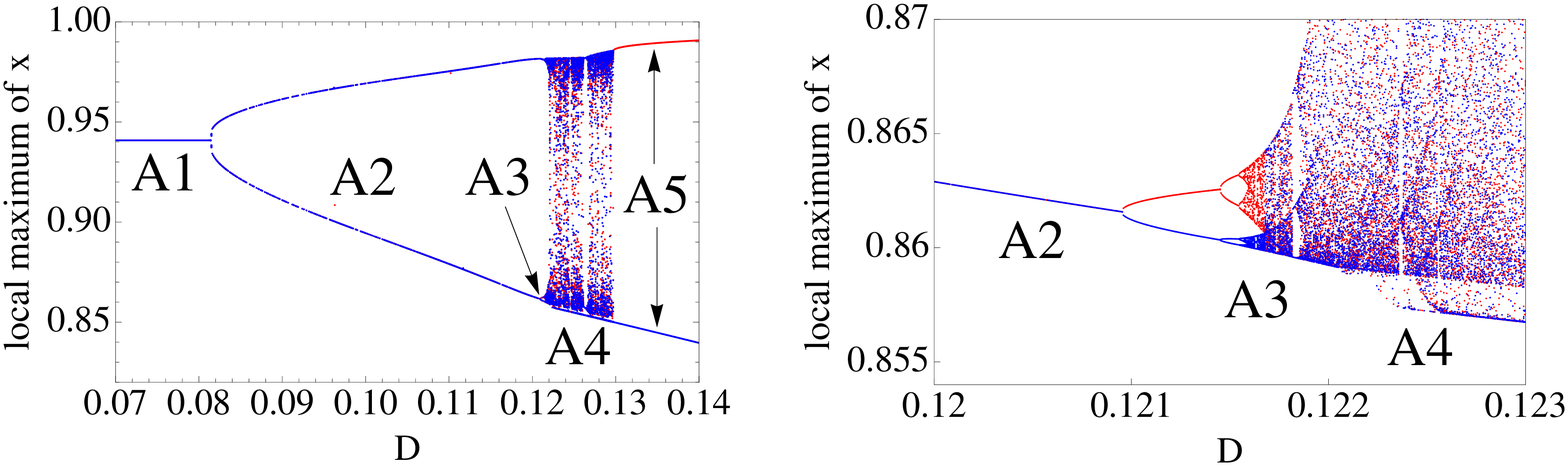}
\caption{Bifurcation diagram of the two-cell dynamics. (a) Local
maxima of $x_i(t)$ for two cells ($i=1,2$; red and blue) plotted at
the steady state (attractor). (b) Details around the A3-A4
transition.
%The difference of color represents the different cells.
As $D$ is increased, the single-cell orbit is destabilized ($A1
\rightarrow A2$), and the two cells take different orbits through
the pitchfork bifurcation (A2). Chaos appears through the
period-doubling cascade (A3) and two chaotic orbits merge so that
the two cells fall into the same chaotic orbit, keeping
desynchronization. With a further increase in $D$, distinct cells
with different oscillations coexist (A5). } \label{Fig233no1}
\end{center}
\end{figure}

Since the desynchronization and differentiation are due to the
interaction, the dynamics of the two cells depends strongly on the
diffusion coefficient $D$. By taking a two-cell system with a fixed
$g_{x,y}$ as above, we can examine the dependence of the two-cell
dynamics on the value of $D$. In Fig. \ref{Fig233no1}, the local
maxima $x_i(t)$ of the two cells $i=1,2$ during the steady state
(either a fixed point or periodic oscillation) are plotted over
time. For small $D$, i.e., for weak cell--cell interaction, the
oscillations of the two cells are synchronized. With an increase in
$D$, the synchronization loses stability, and the oscillations
desynchronize (Fig. \ref{Fig233no1}); coexistence of the two
oscillations occurs for larger $D$.

\begin{figure}[H]
\begin{center}
\includegraphics[width=100mm]{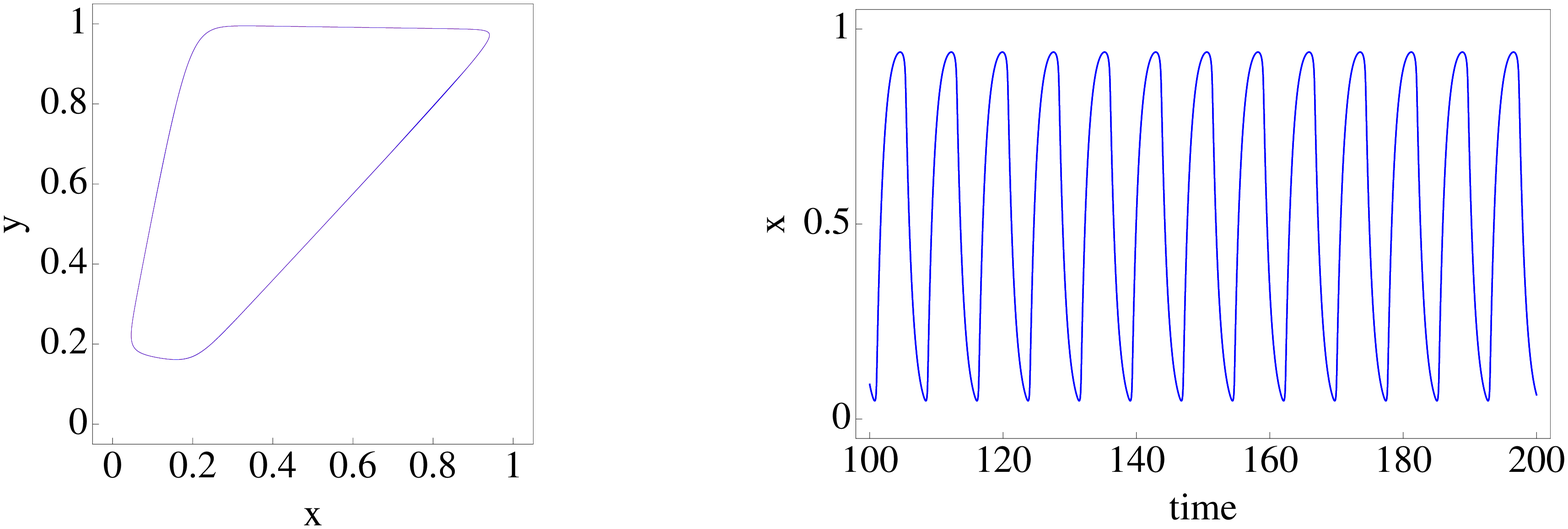}
\includegraphics[width=100mm]{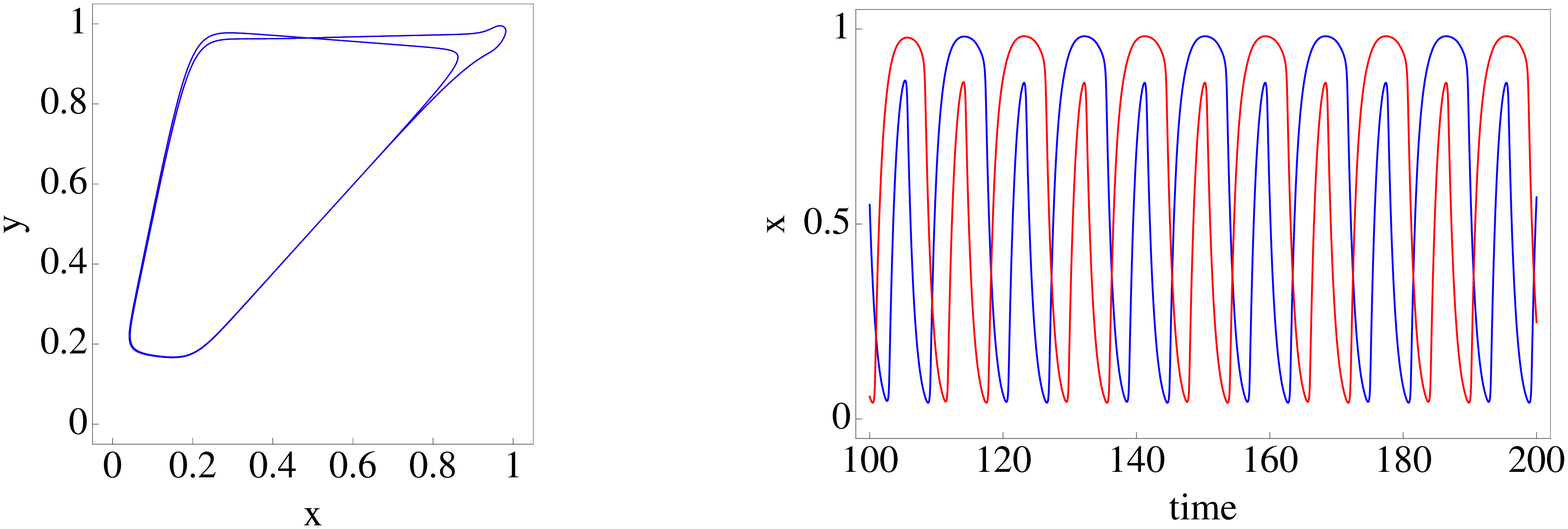}
\includegraphics[width=100mm]{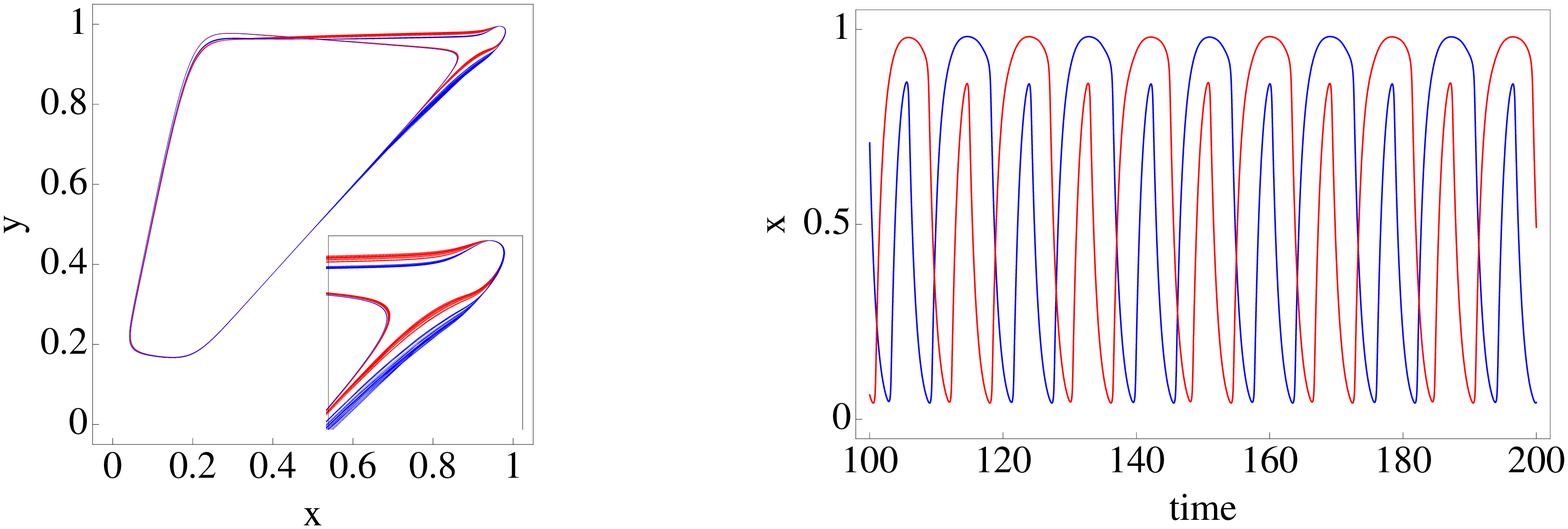}
\includegraphics[width=100mm]{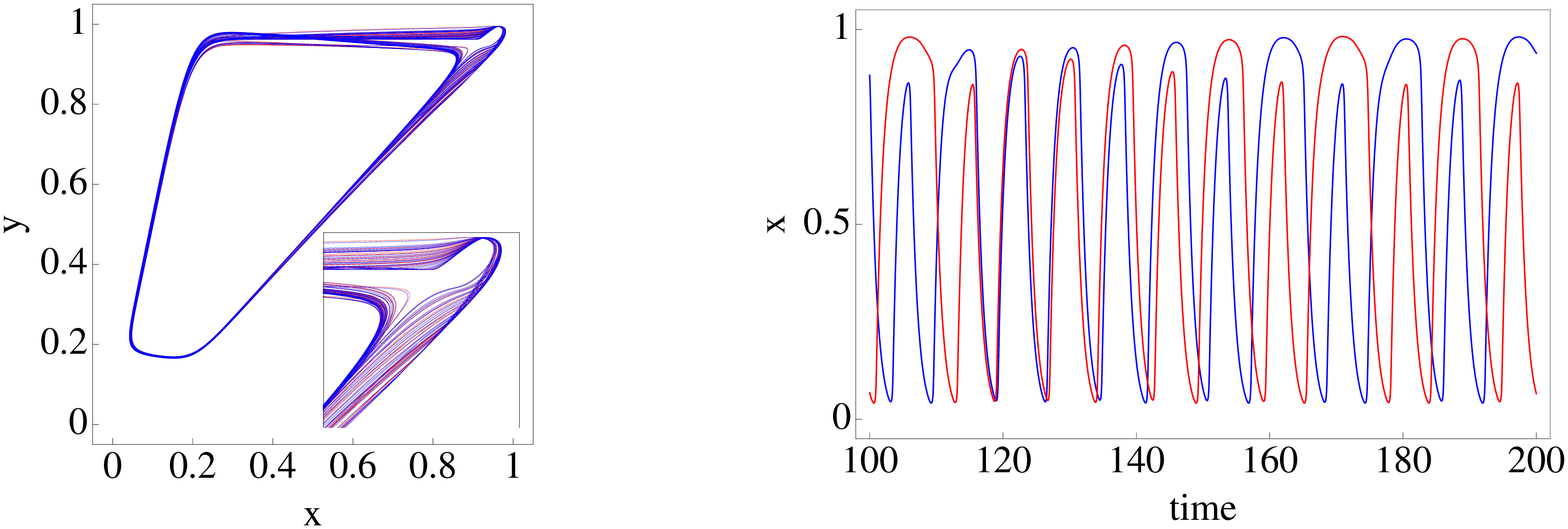}
\includegraphics[width=100mm]{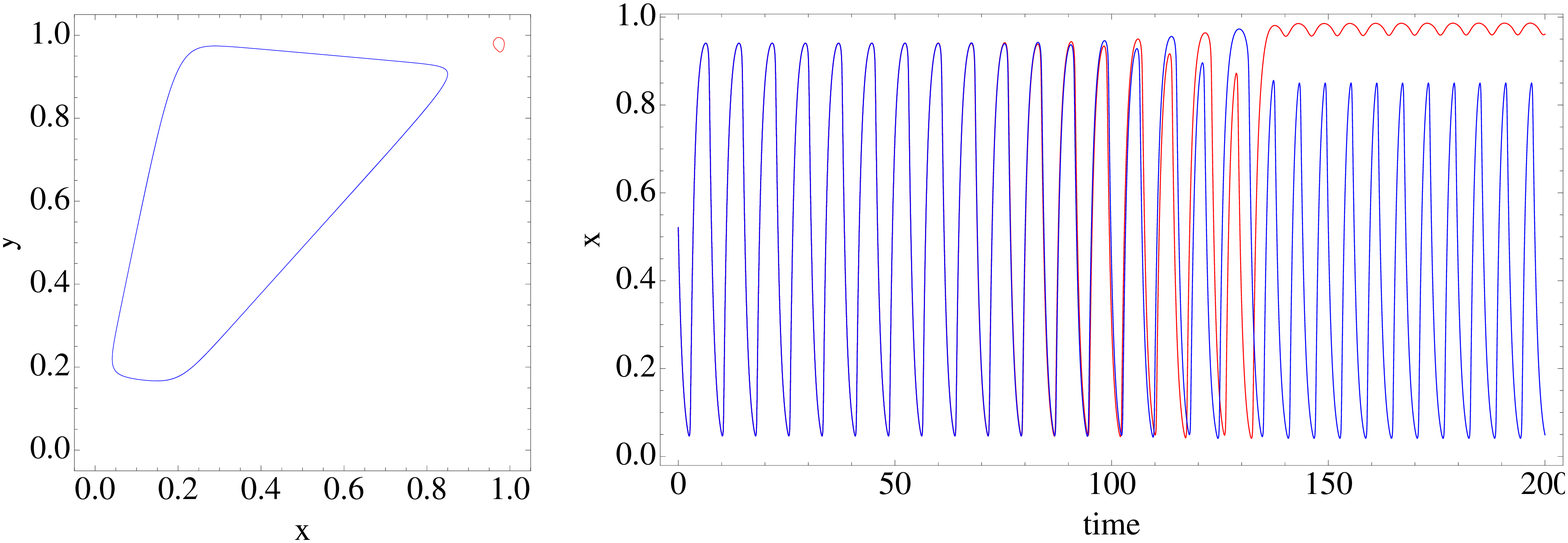}
\caption{Orbit and time series of the two cells. Left: Orbits of
$(x_i(t),y_i(t))$ ($i=1,2$) after the cells reach a steady state
(attractor). Right: The time series of $x_i(t)$. The colors
correspond to the two different cells $i=1,2$. (i) $D=0.0$
corresponding to A1 in Fig. \ref{Fig233no1}. The orbits of the two
cells are synchronized. (ii) $D=0.12$ corresponding to A2. The
orbits are desynchronized. (iii) $D=0.1216$ corresponding to A3. The
orbits of the two cells are desynchronized and take different loci.
(iv) $D=0.1225$ corresponding to A4. The two cells show chaotic
dynamics and are desynchronized, while as a whole they are on the
same attractor. (v) $D=0.132$ corresponding to A5. After
desynchronization, the dynamics of the two cells are split into two
distinct behaviors. The differentiation discussed in section III.B.2
corresponds to this region, while $D$ adopted in that discussion was
0.14. } \label{Fig233gno1}
\end{center}
\end{figure}

The corresponding changes in the time series and trajectories in
$(x,y)$ space are shown for each region (A1, A2, ..., A5) in Fig.
\ref{Fig233gno1}. From A1 to A2, a period-doubling bifurcation causes the
oscillations of the two cells to be out of phase. 
From A2 to A3 a pitchfork bifurcation occurs. 
Then, the period-doubling cascade to chaos appears from A3 to A4, 
where the two oscillations are chaotic with desynchronization. 
At the transition from A4 to A5, chaotic orbit touches with saddle, and crisis occurs.
 The chaotic orbit in the four dimensional space with two cells becomes unstable and is replaced by two stable periodic orbits.
  Each of the oscillations is now periodic; one with a large amplitude close to the original limit cycle and the other with
a tiny amplitude near the fixed point.
 Form the view point of single cell dynamics, it can be regarded as occurrence of SNIC bifurcation,
  when we regard diffusion term as time dependent parameter.
 In region A5, the asymmetric differentiation from the stem-type cell, which was discussed earlier, occurs.
We also note that the bifurcation  in the case of oscillation death follows almost the same sequence as the present case.

\subsection{Robustness in cell number regulation}

\begin{figure}[H]
\begin{center}
\includegraphics[width=80mm]{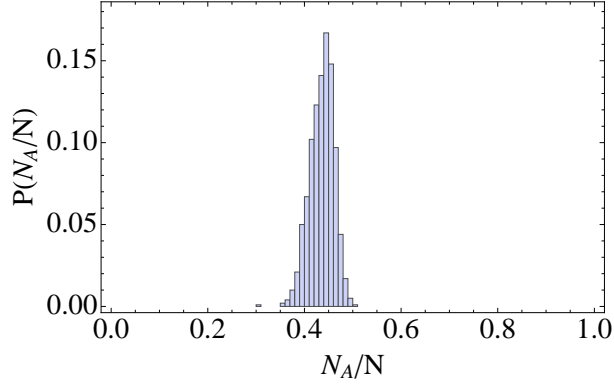}
\caption{Histogram of the ratio of the number of original type-$A$
cells with a large-amplitude limit cycle to the total number of
cells. Here, $N=100$ cells are used with 200 sets of randomly chosen
initial values for $(x_i,y_i)$ distributed homogeneously between 0
and 1. After the expressions reached the steady state, the ratio is
computed to construct the histogram.} \label{Fig234no1}
\end{center}
\end{figure}

\begin{figure}[H]
\begin{center}
\includegraphics[width=80mm]{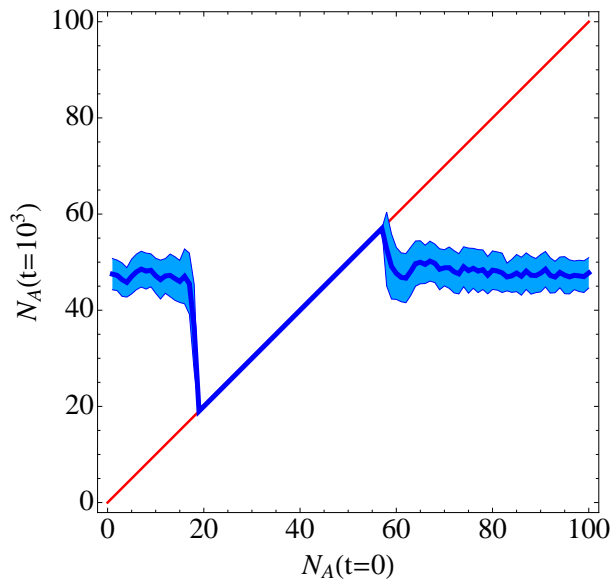}
\caption{The range of differentiated cell types obtained from an
initial condition of two types of cells. The abscissa is the initial
number of type A cells (the original cell type with a
large-amplitude limit cycle) among 100 cells, and the ordinate is
the number of such cells after reaching the steady state at $t=100$.
For each initial number of type A cells, 200 initial conditions are
chosen for the $(x_i,y_i)$ cells. The dark blue line shows the
average of the ratio of type A cells over the 200 samples, while the
blue bars show the range of the standard deviation from the
distribution. } \label{Fig234no2}
\end{center}
\end{figure}

We have identified three classes of cell differentiation in terms of
dynamical systems theory. The first class is that in which multiple
attractors exist, where the switch among the attractors by noise
leads to cell differentiation. The second is a Turing-type class in
which the initial cell state is unstable and leads to two different
states. The third class is asymmetric differentiation from an
oscillatory state. Only the third class has a type of cell that can
both proliferate and differentiate, which reflects the nature of
stem cells.

In the multiple-attractor case, the differentiation is due to noise
so that the time course of the development is stochastic; the number
ratio of each cell type is unregulated. In the Turing-type case, the
original cell type just disappears, so that there is no
stem-cell-like cell.

One merit of asymmetric differentiation lies in both the existence
of stem-cell-type cells and the robustness of the ratio of each
cell-type number against noise \cite{Furusawa-KK01,SFK} or a change
in the initial conditions. This robustness is expected, as the
differentiated cell type appears as a result of the instability of
the state consisting of the first cell type only, and the two cell
types stabilize each other through cell--cell interactions.
%%%%%%%%%%%%%%%%%%%%323-added
In this section, we call the original cells with a large-amplitude limit cycle as type-A cells, 
and differentiated cells with a small-amplitude limit cycle as type-B cells and study the robustness in the number ratio of type A and B cells.
%%%%%%%%%%%%%%%%%%%%%%%%
To examine this robustness, we first simulated our model starting
with a single cell for a given initial condition and progressively
added noise. We confirmed that the ratio of one type of cell to the
total number of cells has a narrow distribution even in the presence
of noise, once the total number of cells reached a particular number
(e.g., 32).

We then started with $N$ initial cells with random initial values of
$(x_i,y_i)$ distributed homogeneously between 0 and 1. As shown in
Fig. \ref{Fig234no1}, the ratio has a narrow distribution centered
around 0.43. To check the possible range of numbers of the two cell
types, we simulated the model with initial conditions of the
expression levels so that there were $N_A$ type-$A$ cells and
$N-N_A$ differentiated cells. If $N_A$ is initially too large, then
the state is unstable, and some of the $N_A$ cells show a SNIC
bifurcation to lose the oscillation and differentiate to type-$B$
cells with fixed-point behavior. If $N_A$ is initially too small,
(i.e., the proportion of differentiated cells of type $N_B$ is too
large), then the type-$B$ state is unstable and some of these
type-$B$ cells regain the oscillation to de-differentiate to
type-$A$ cells. Thus, there are upper and lower bounds on the ratio
of the number of type-$A$ cells. For example, in the parameter
values used in Fig. \ref{Fig234no2}, this range is 0.2 to 0.55.
Therefore, the robustness of the two types of cells is a consequence
of the present asymmetric differentiation based on the SNIC.

\subsection{Complex cell differentiation by combining two-gene motifs}

\begin{figure}[H]
\begin{center}
\includegraphics[width=50mm]{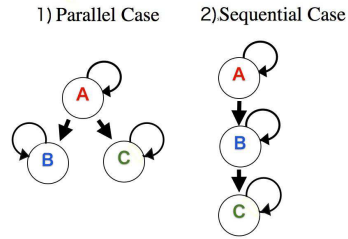}
\caption{Basic cell lineage diagram. Complex differentiation is
composed of a combination of these two processes. (a) Parallel case:
two types of cells are created from one type of cell. (b) Sequential
case: cells differentiate in series. } \label{Fig24no1}
\end{center}
\end{figure}

\begin{figure}[H]
\begin{center}
\includegraphics[width=100mm]{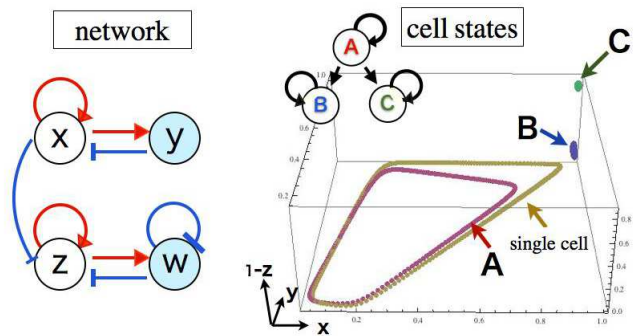}
\caption{The parallel case. (a) The GRN. (b) The dynamics of the
three cell types are plotted as orbits in $x,y,z$ space. When
asymmetric differentiation occurs in the $x$--$y$ network, the
protein $z$ is expressed, which leads to a Turing instability. As a
result, cells with small-amplitude oscillations in the $x$--$y$
plane split into two states with regard to the value of $z$
(type-$B$ and -$C$ cells).The parameter values that show this differentiation are$(D_{y},D_{w},g_{x},g_{y},g_{z},g_{w})=(0.2,0.2,-0.1,0.2,0.998,0)$
} \label{Fig24no2}
\end{center}
\end{figure}

\begin{figure}[H]
\begin{center}
\includegraphics[width=100mm]{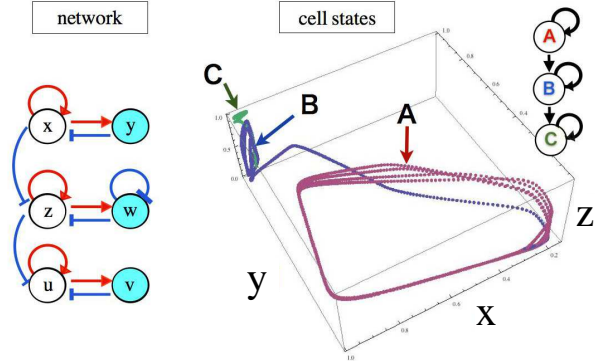}
\caption{The sequential case. (a) The GRN. (b) The dynamics of the
three cell types are plotted as orbits in $x,y,u$ space. Asymmetric
differentiation occurs in the $x$--$y$ network and some cells change
from type-$A$ to type-$B$ cells. The type-$B$ cells show asymmetric
differentiation in the $u$--$v$ gene network, which results in some
type-$B$ cells changing into type-$C$ cells.
The parameter values that show this differentiation are $(D_{y},D_{w},D_{v})$$=(0.2,0.2,0.2)$,  
$(g_{x}$,$g_{y}$,$g_{z}$,$g_{w}$,$g_{u}$,$g_{v}$)
=(-0.1,0.2,0.99,0.0,-0.07,-0.2).$J_{m\ell}=1$ if the gene $\ell$ activates the expression of $m$,
$-1$ if it inhibits the expression, and 0 if there is no regulation. But note that $J_{uz}=-0.07$.
 } \label{Fig24no3}
\end{center}
\end{figure}

Most of the present cells consist of a large number of cell types
that are generated through successive differentiations. The entire
cell lineage can be constructed by combining the following two
differentiation processes.

(i) The parallel case: bifurcation into two types of cells from one
type of cell, as shown in Fig. \ref{Fig24no1}(a).

(ii) The sequential case: cells differentiate in series to form a
hierarchical differentiation, %Type-$B$ cells differentiate from
%type-$A$ cells and type-$C$ from type-$B$,
as shown in Fig.~\ref{Fig24no1}(b).

Complex cell differentiations as observed in a hematopoietic system,
for example, are shaped by combining these parallel and sequential
differentiations. We show here that these two basic forms are
designed straightforwardly by combining the Turing-type module and
the asymmetric differentiation module.

The parallel case can be achieved by a gene network as shown in Fig.
\ref{Fig24no2}. The Turing-type module is regulated by the
asymmetric differentiation module. There are four proteins $x,y,z,$
and $w$ in total whose concentrations are represented by the
corresponding variables. The protein levels of the cells initially
show oscillations, forming a large-amplitude limit cycle in the
$x$--$y$ plane (type-$A$ cells). As the number of cells increases,
asymmetric differentiation occurs in the $x$--$y$ plane to form a
state with constant, higher expressions of $x$ and $y$ via the
mechanism mentioned earlier. With this activation of $x$, a
Turing-type bifurcation in the $z$ and $w$ protein expressions is
induced. Both $z$ and $w$ are expressed in the beginning, but
following the suppression of $z$, the state differentiates into two
types of cells with higher and lower expressions of both $z$ and $w$
based on the Turing-type mechanism. Hence, the differentiation from
stem-cell-type $A$ to two types of cells $B$ and $C$ occurs, as
shown in Fig. \ref{Fig24no2}.

The sequential case, however, involves the gene network shown in
Fig. \ref{Fig24no3} in which two asymmetric differentiation motifs
($(x,y)$ and $(u,v)$) are connected in sequence with the intervening
Turing motif ($z,w$). Here again, all the cells initially show
oscillation and form a large-amplitude limit cycle in the $(x,y)$
plane (type-$A$ cells). As the number of cells increases, asymmetric
differentiations occur and some cells differentiate into type-$B$
cells, which fall on a small-amplitude limit cycle in the $(x,y)$
plane. After this bifurcation, the expression of $x$ is constantly
activated in the type-$B$ cells, which then suppresses the expression
of $z$ and activate the expression of $u$, triggering the asymmetric
differentiation of the expressions of ($u,v)$ and leading to the
differentiation to type-$C$ cells with a constant activation of $u$.
Thus, hierarchical, sequential differentiation from type-$A$ to $B$
and then to $C$ is generated, through which the oscillation
amplitude decreases accordingly. (Here, the intermediate Turing
module is used to suppress the expression level of $z$ when the
expression of $x$ is constantly activated and to activate the
expression otherwise. This is not an essential component, as we
believe other network forms can be adopted.)

\section{Discussion}

We have studied in this paper an interacting cell model consisting
of two genes (protein expressions) and extracted, from extensive
simulation, minimal gene networks that show differentiations. These
differentiations were classified as Turing, oscillation death, or
asymmetric differentiation with remnant oscillations (see also
\cite{Volkov,Volkov2} for other possible types for differentiation
including an inhomogeneous limit cycle in which two limit cycles
coexist). Only asymmetric differentiation was shown to allow for
cells with stemness, i.e., compatibility with both proliferation and
differentiation. The differentiation is understood as a SNIC bifurcation which
is triggered by cell--cell interactions with the cell
number of the original cell type being the bifurcation parameter. It
was shown that each cellular state is stabilized according to the
cell--cell interaction, which depends on the number distribution of
each cell type. The number ratio is thus regulated autonomously. In
this sense, the effective bifurcation parameter due to cell--cell
interactions is ``self-consistently'' determined \cite{Nakajima}. A
theoretical analysis for such a self-consistent bifurcation should
be developed in future.

With the present study, we can now answer the questions addressed in the Introduction. 
(i) Which attractor describes the two conflicting functions in stem cells,
i.e., proliferation and differentiation?  -- A single-cell attractor providing
stemness is a limit-cycle that is close to the point of SNIC bifurcation. The limit-cycle attractor
provides stability, while the cellular state is easily switched to a different state
with the aid of SNIC bifurcation, due to cell-cell interaction. (ii) How are initial
conditions for different attracting states selected through the
course of development?  -- With the cell-cell interaction, desynchronization of oscillations
follows, which diversifies the cellular states.  Then, with the cell-cell interaction,
states of some cells are kicked out from the original basin of attraction, triggered by SNIC bifurcation.
 (iii) How is the stability of the developmental course explained,
which possibly includes regulation of the cell--cell interactions?
-- Since the desnchronization and bifurcation occur as a result of the increase in cell number, the
differentiation timing is almost deterministic through the developmental course.  The cell types thus generated
are stabilized with each other through cell-cell interaction, which depends on the number ratio of each cell type.
Hence the ratio of each cell-type is regulated so that it stays within a certain proportion, and it is robust to noise.  

We were able to extract the minimal gene expression network for the
stemness and demonstrated that it consists of two genes; one
activates and suppresses the other, while the other has an
activation path to itself. Due to the simplicity, the network can
work as a motif \cite{Alon} for complex cell differentiation in
general. In fact, several networks previously studied for three or
more genes \cite{SFK} include the present minimal network motif as
their core component. By including feedback or feed-forward path(s)
for gene expression networks to the extracted minimal network,
stem-cell-type behavior as well as the differentiation process
further enhances the robustness against change in the parameter
values.

Although we have adopted a simple form of the threshold expression
dynamics, the bifurcation analysis developed here shows that the
present mechanism for differentiation is possible in other forms of
the expression dynamics, as long as the SNIC occurs due to the
cell--cell interaction or signal molecule. For example, by using the
Hill form for the expression,

\begin{equation}
\begin{split}
\epsilon\frac{dx_i(t)}{dt}&=\frac{(\frac{x_i(t)}{K_{x_1}})^{n_1}}{1+(\frac{x_i(t)}{K_{x_1}})^{n_1}}\frac{1}{1+(\frac{y_i(t)}{K_y})^{n_2} }-x_i(t)+I_1\\
\frac{dy_i(t)}{dt}&=\frac{(\frac{x_i(t)}{K_{x_2}})^{n_3}}{1+(\frac{x_i(t)}{K_{x_2}})^{n_3}}-y_i(t)+I_2+D\Bigl(\sum_{k=1}^Ny_k(t)-y_i(t)\Bigl),\\
\end{split}
\end{equation}
the present differentiation progresses for appropriate values of the
parameters (with Hill coefficients $n_i$ ($i=1,2,3$) of sufficiently
large values).

%$K_{x_1}=0.62,\;K_{x_2}=0.65,\;K_y=0.5,n_1=3, n_2=5,n_3=5$\\ $I_1=0.2,\;I_2=0.3,\epsilon=0.125,D=0.5$\\

Real GRNs, however, include more than a thousand genes, and actual
networks are quite complicated. In spite of this, the present
network motif, as it is so small, can be easily included in the
networks of the present cell. A combination of the present two-gene
network motifs can lead to differentiations of a complex cell
lineage, as was demonstrated here. It will be important to extract
such network motif combinations in relation to the observed complex
differentiation \cite{real}.

Acknowledgement:

The authors would like to thank Shuji Ishihara, Chikara Furusawa, Benjamin Pfeuty,
and Narito Suzuki for useful discussions. This work was partially
supported by a Grant-in-Aid for Scientific Research (No. 21120004)
on Innovative Areas ``Neural creativity for communication'' (No.
4103) and the Platform for Dynamic Approaches to Living System from
MEXT, Japan.

{}
\end{document}